# A conventional positron source for International Linear Collider

Tsunehiko Omori[a]*, Tohru Takahashi[b], Sabine Riemann[c], Wei Gai[d], Jie Gao[e],

Shin-ichi Kawada[b], Wanming Liu[d], Natsuki Okuda[f], Guoxi Pei[e],

Junji Urakawa[a], and Andriy Ushakov[g]

[a]KEK: High Energy Acceleratoor Research Organization, 1-1 Oho, Tsukuba-shi, Ibaraki 3050801, Japan
[b]Graduate School of Advanced Sciences of Matter, Hiroshima University, 1-3-1 Kagamiyama, Higashi-Hiroshima, 739-8530, Japan
[c]Deutsches Elektronen-Synchrotron, DESY, Platanenallee 6, D-15738 Zeuthen, Germany
[d]Argonne National Laboratory, 9700 S. Cass Avenue Argonne, IL 60439, USA
[e]Institute of High Energy Physics, 19B YuquanLu, Shijingshan District, Beijing, 100049, China
[f]Department of Physics, Graduate School of Science, The University of Tokyo, 7-3-1 Hongo, Bunkyo-ku, Tokyo 113-0033, Japan
[g]University of Hamburg, Luruper Chaussee 149, D-22607 Hamburg, Germany

* Corresponding author. Tel.:+81-29-8645370; fax: +81-29-8642580
  E-mail address: tsunehiko.omori@kek.jp

## Abstract

A possible solution to realize a conventional positron source driven by a several-GeV electron beam for the International Linear Collider is proposed. A 300 Hz electron linac is employed to create positrons with stretching pulse length in order to cure target thermal load. ILC requires about 2600 bunches in a train which pulse length is 1 ms. Each pulse of the 300 Hz linac creates about 130 bunches, then 2600 bunches are created in 63 ms. Optimized parameters such as drive beam energy, beam size, and target thickness, are discussed assuming a L-band capture system to maximize the capture efficiency and to mitigate the target thermal load. A slow rotating tungsten disk is employed as positron generation target.

## 1. Introduction

The International Linear Collier (ILC)[1] is an electron positron linear collider project which employs the superconducting RF acceleration technology in the main linacs. This allows to accelerate high current beams with pulses of 1ms duration consisting of about 2600 bunches of positrons (electrons), each bunch contains $2 \times 10^{10}$ positrons (electrons). Such high currents are required to realize the high luminosity, $2 \times 10^{34}$ cm$^{-2}$s$^{-1}$ at $E_{CM}$ = 500 GeV, for the

ILC.

However, it is a very challenging issue to design the source for such a high current positron beam. One of biggest risk areas is the thermal load on the positron production target.

The baseline choice of the ILC positron source is the helical undulator scheme. After accelerating the electron beam in the main linac, it passes a 150 m long helical undulator to create a circularly polarized photon beam, and goes to the interaction point [2]. The photon beam hits the production target and generates electron-positron pairs. The positrons are captured, accelerated to 5 GeV, damped, and then accelerated to the collision energy in the main linac. Thus the undulator based positron generation gives interconnection to nearly all sub-systems of the ILC.

Designing and operation of such a large-scale interconnected system is a challenge and strict constraints are given to the positron source and the target heat load by the time structure of the beams. We are constrained to create 2600 bunches in 1ms, for example. However, it is one advantage of the helical undulator scheme that it provides a polarized positron beam. This advantage will be essential when the LHC will find physics scenarios which can be studied with higher sensitivity if both beams of a electrons-positrons collider are polarized[3].

On the other hand, it could be that the LHC measurements give strong hints to physics scenarios where positron polarization would add only marginal information. Then a conventional unpolarized positron source could be sufficient to reach the physics goal of the ILC. The conventional source suggested in this paper would reduce the thermal load in the positron target by distributing the thermal exposure over the time and using an electron beam with a large cross section. So the risk for the target area can be minimized.

Up to now, only the conventional positron generation scheme has been experienced in real accelerators [4]. With these experiences we are able to control risks in a limited area, that is the target area. The potential risks of the conventional source system are known and could be minimized if used also for the ILC. However, if polarized positrons are required in a later stage of the experiments, the positron source has to be replaced.

The proposed ILC positron source contains risks only in the target area. Therefore, we concentrate to cure these risks in two ways: (1) pulse stretching by 300 Hz generation; the proposed scheme creates 2600 bunches in about 60 ms, and (2) optimized drive beam and target thickness parameters.

## 2. Pulse stretching by 300 Hz generation

The proposed positron source eases the target risk by stretching the pulse length of a bunch-train. The ILC repetition rate, 5 Hz, is rather slow. We have an interval of 200 ms between two pulses. This gives enough time for pulse stretching. A 300 Hz electron linac is employed to create positrons. Since the ILC requires about 2600 bunches in a train of pulse length 1ms, each pulse of the 300 Hz linac must create about 130 bunches to achieve 2600 bunches in 60 ms.

The positrons created at the source are sent to the damping ring to make the emittance small. In designing the conventional positron source, the authors choose parameters such that they fulfill to requirements of the ILC baseline damping ring. The bunch-to-bunch separation in the damping ring is 6.15 ns, therefore the bunch-to-bunch separation of 6.15 ns in the 300 Hz linac is chosen. In order to avoid instabilities in the damping ring caused by electron clouds, the positron beam has a mini-train structure where about 40 bunches form a mini-train. The gaps of about 100 ns between adjacent mini-trains are the key to prevent instabilities.

The beam structure in the 300 Hz linac is similar. One RF pulse accelerates three mini-trains with inter-mini-train gaps. This package of three mini-trains is named triplet in this article. Each mini-train contains 44 bunches.

In this article, we assume the ILC nominal beam parameters, but with slight modifications. In the original ILC nominal parameter set one RF pulse in the main linac has 2625 bunches. Here, we assume 2640 bunches per pulse.

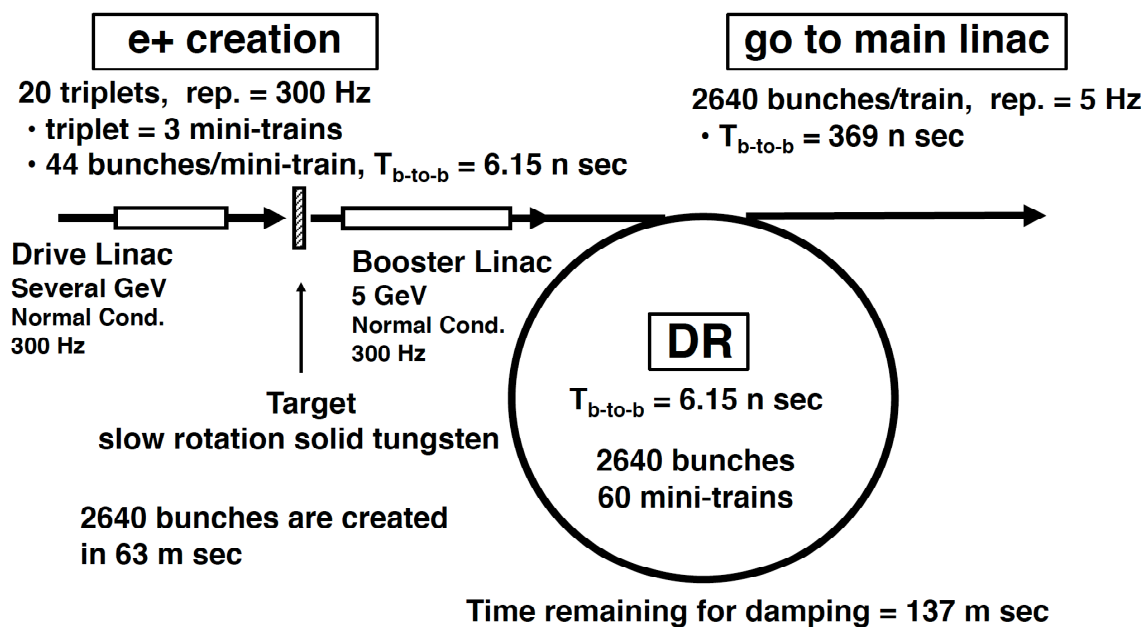

Figure 1. Schematic view of the 300 Hz scheme.

Figure 1 shows the schematic view of the 300 Hz scheme. In the ILC, we can employ different bunch structures and pulse structures in the positron source, in the DR, and in the main linac. In the 300 Hz scheme, we employ the triplet mini-train structure in the positron source (Figure 2) and the mini-train structure in the DR. With 6.15 ns bunch-to-bunch separation and 44 bunches per mini-train, the length of a mini-train is 264 ns. Since a triplet contains three mini-trains, it consists of 132 bunches and we need 20 triplets to form 2640 bunches.

There are gaps of about 100 ns between the mini-trains in a triplet. This triplet structure is required in order to match the bunch timing structure to the fill pattern of DR. Since the triplets are produced by the electron beam they have a repetition rate of 300 Hz. Therefore, we have 3.3 ms between two triplets.

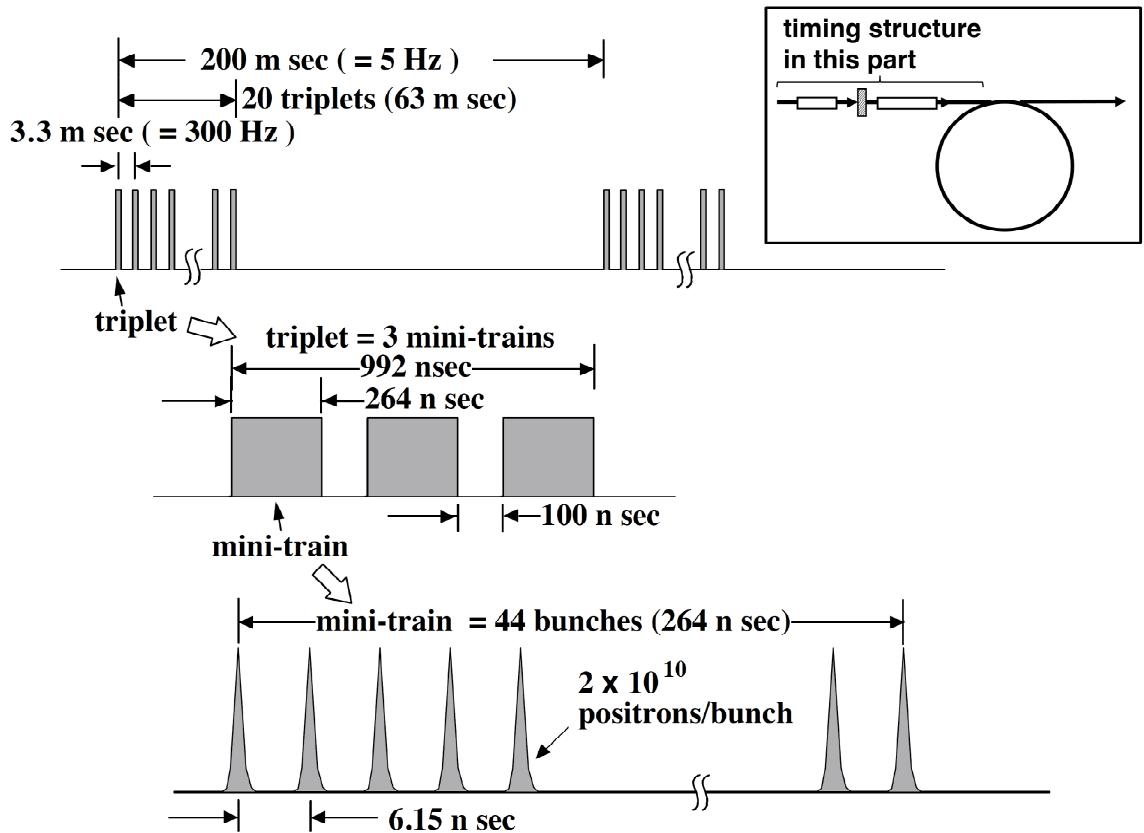

Figure 2. Timing structure in the positron source and in the booster linac.

We assume to employ a rotating target made of tungsten alloy. Due to the target rotation, different triplets hit different parts of the target, and the target does not need to survive the impact of 2640 bunches at the same spot. Figure 3 shows the results of simulations of the instantaneous temperature rise caused by successive triplets impinging on the target; the tangential speeds assumed are (a) 5 m/s, (b) 2 m/s, (c) 1 m/s, and (d) 0.5 m/s. A drive beam energy of 6 GeV and a bunch charge of 3.2 nC are assumed, the spot size of the drive beam at the target is 4 mm (rms) and the target thickness is 14mm. The choice of the drive beam parameters is discussed later.

As shown in Figure 3d, the temperature rise of the target is about 1300 K when we choose 0.5 m/s. Since the melting point of tungsten is 3697 K, a tangential speed of 0.5 m/s is sufficient to tolerate the heat load. The rotation speeds assumed in this article are much less than 100 m/s which is assumed in ILC baseline design. This is one advantage of the 300 Hz scheme. In the simulation, we did not consider cooling of the target. The analysis of the equilibrium temperature for a realistic target design is a future plan.

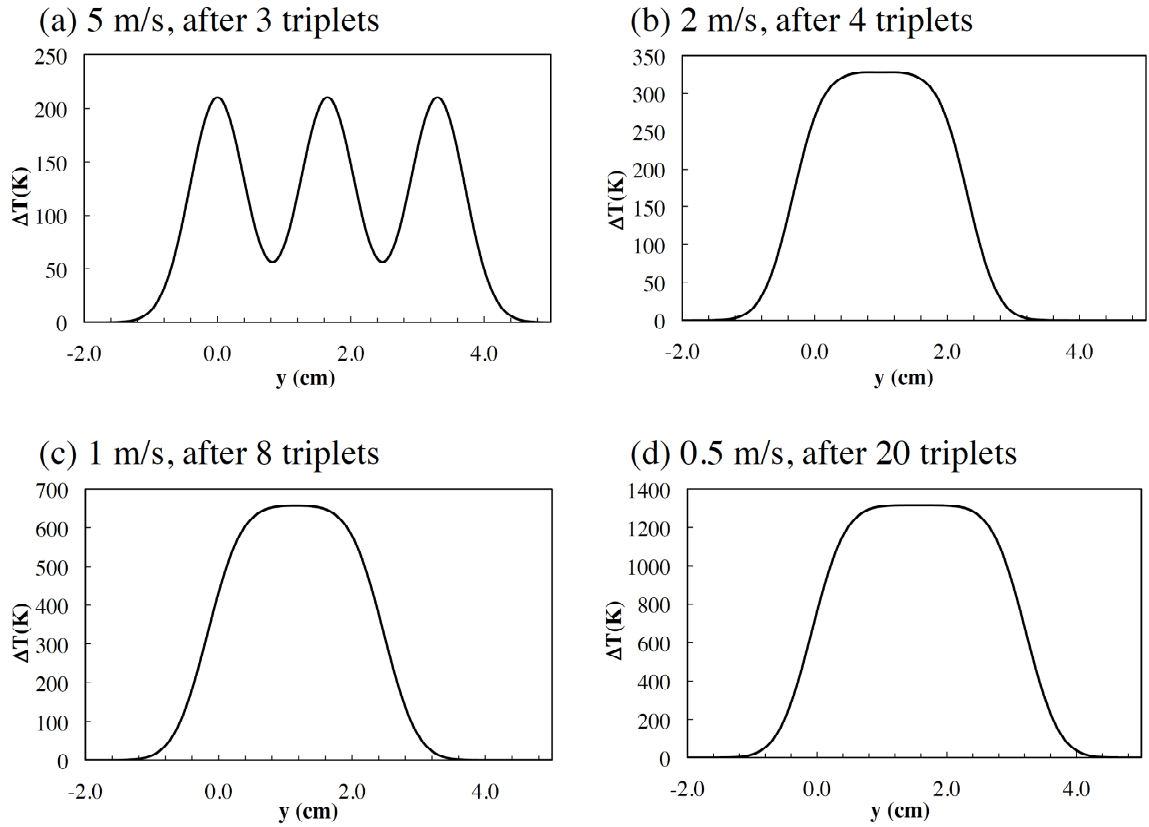

Figure 3. Instantaneous temperature rise of the target hit by successive triplets. The tangential speeds are (a) 5 m/s, (b) 2 m/s, (c) 1 m/s, and (d) 0.5 m/s. In the simulation, a drive beam energy of 6 GeV and a bunch charge of 3.2 nC are assumed.

A flux concentrator is assumed as Adiabatic Matching Device in the capture section. In the 300 Hz scheme, the required pulse length of the flux concentrator is short; it is about 1 μs. This is the same as that of existing flux concentrators, so the technology is available. Further, the short pulse length allows us to use a high acceleration gradient. The details of the capture system are discussed later.

After the target and capture device, the positron energy is boosted to 5 GeV in a 300 Hz normal conducting linac. A kicker with pulse length of about 1 μs and repetition rate 300 Hz is employd to send the positrons to the DR. One kicker pulse sends a triplet to DR. Also the kicker with 1 μs pulse length can be build with existing technology.

After the damping, bunches are extracted from the DR by a fast kicker, sent to the bunch compressor, and then to the main linac. This part remains the same as in the ILC baseline design with an undulator positron source, the bunch-to-bunch separation is 369 ns after the extraction (Figure 4).

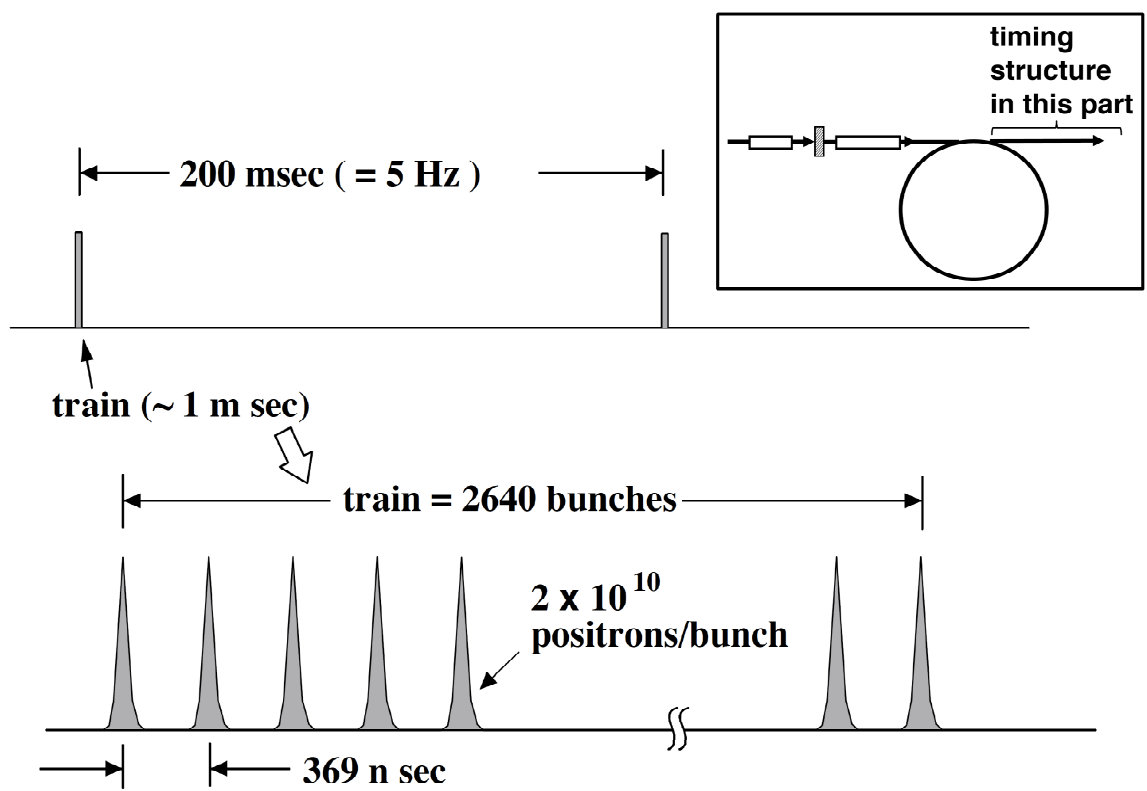

Figure 4: Time structure after the damping ring.

## 3. Optimization of drive beam and target parameters.

In order to optimize the parameters for the conventional source, positron yield, peak energy deposit density (PEDD) and total energy deposit in the target were calculated for various combinations of drive beam energy and target thickness. Since PEDD and capture efficiency depend on the transverse size of the drive beam, estimations were also performed for several drive beam sizes. We used Geant4 for the simulation of positron generation in the target and particle tracking in the subsequent capture section.

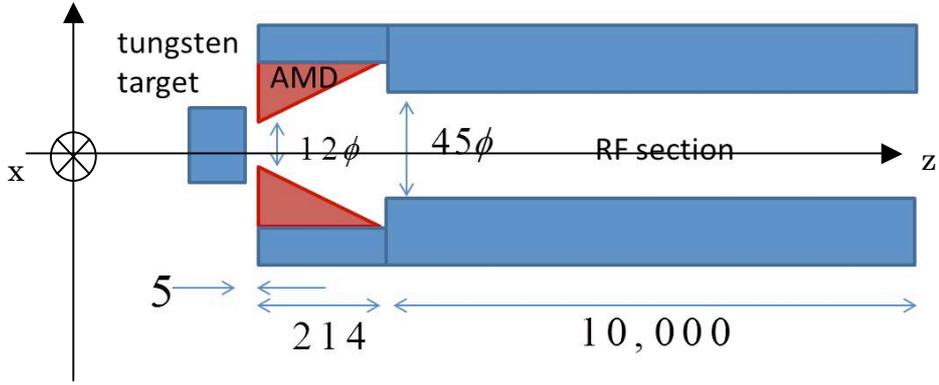

Figure 5. Layout of the target and positron capture section implemented in the simulation.

Figure 5 shows the target and the capture section assumed in the analysis. The target material was solid tungsten. An adiabatic matching device (AMD) was assumed as optical matching followed by a L-band pre-accelerator. The longitudinal magnetic field of the AMD is described as

$$Bz(z-z_0) = \frac{B_0}{1+\mu(z-z_0)} + Bsol$$

where z = 0 is the back end of the tungsten target. The maximum field, $B_0$, was 7 T and the taper parameter μ was 60.1 m$^{-1}$. The parameter $z_0$, a gap between the target and the AMD field, was assumed 5 mm to accommodate the rotation target. The input aperture of the AMD was 12 mm in diameter. The radial field is calculated according to the prescription described in Ref [5] as

$$Br(r,z) = -\frac{1}{2}r\frac{\partial B(z)}{\partial z} + \frac{1}{16}r^3\frac{\partial^3 B(z)}{\partial z^3},$$

where r is the transverse distance to the center of the AMD and of the RF section. After the AMD, a RF accelerating field of 1.3 GHz traveling plane wave with the amplitude of 25 MV/m was applied from z = 219 to 10219 mm. The aperture of the RF section was assumed to be 45 mm in diameter. In addition to the AMD field, a constant magnetic field $Bsol = 0.5$ T was applied in the AMD and RF section, i.e., from z = 5 to 10129 mm. To calculate the energy deposit in the AMD and RF structures, the outside of the applied fields in lateral direction, indicated by solid areas in Figure 5, was assumed to be iron while the inner area was air of 10$^{-7}$ Pa to mimic the inside of the accelerator structure.

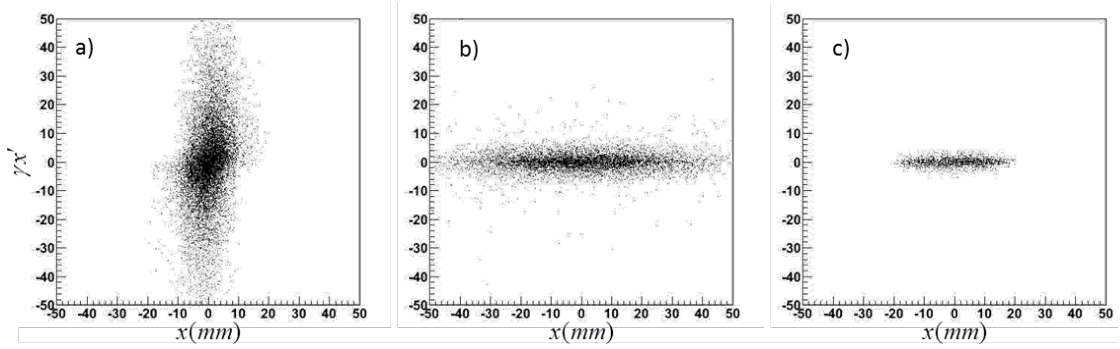

Figure 6. Transverse phase space distribution of positrons at the entrance of the AMD a), the exit of the AMD b) and the exit of the RF section c). A drive beam energy of 6 GeV and a beam spot size of 4 mm (rms) on the target are assumed.

Simulations were performed by changing the target thickness and the energy of the drive electron beam for beam sizes σ = 2.5 mm and 4.0 mm. Figures 6a, 6b and 6c show typical distributions of positrons in the transverse phase space at the input (z = 5 mm) and output (z = 214 mm) of the AMD and at the exit of the RF section (z = 10,219 mm). After the tracking simulation, the normalized emittance for Figure 6c is calculated as

$$\gamma \varepsilon_x = \sqrt{\langle x^2 \rangle \langle (\gamma x')^2 \rangle - \langle x(\gamma x') \rangle} .$$

The twiss parameters for the distribution were obtained:

$$\alpha_x = -\langle x(\gamma x') \rangle / \varepsilon_x$$
$$\beta_x = -\langle x^2 \rangle / \varepsilon_x$$
$$\gamma_x = (1+\alpha_x^2)/\beta_x$$

Then, the parameter $A_x \equiv \gamma_x x + 2\alpha_x x(\gamma x') + \beta_x (\gamma x')^2$ was calculated for each particle, as well $A_y$ for the y direction. The transverse acceptance of the damping ring was constraint to $A_x + A_y < 0.09 mm$. Prior to the estimation of the accepted number of positrons, the phase of the RF field was scanned to maximize the number of positrons.

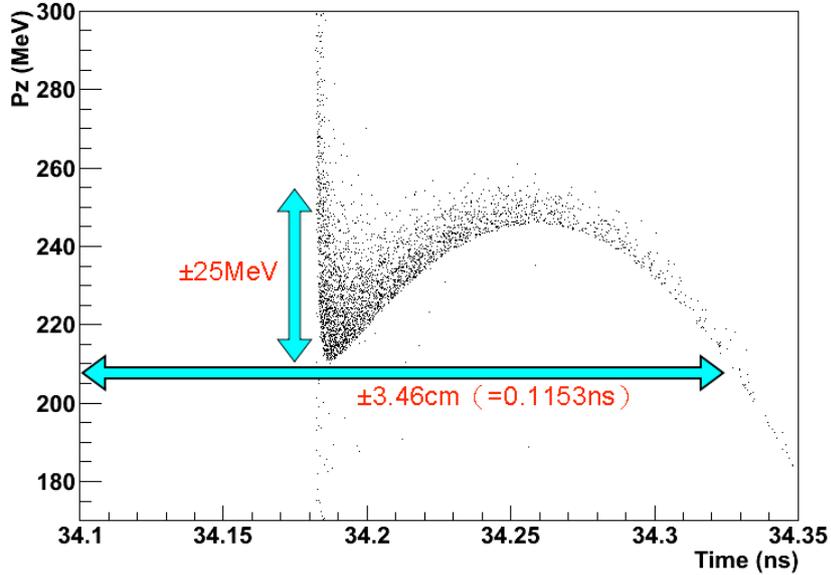

Figure 7. Longitudinal phase space distribution at the exit of the RF section. The acceptance of the damping ring is indicated by arrows. A drive beam energy of 6 GeV and a beam spot size of 4 mm (rms) on target are assumed.

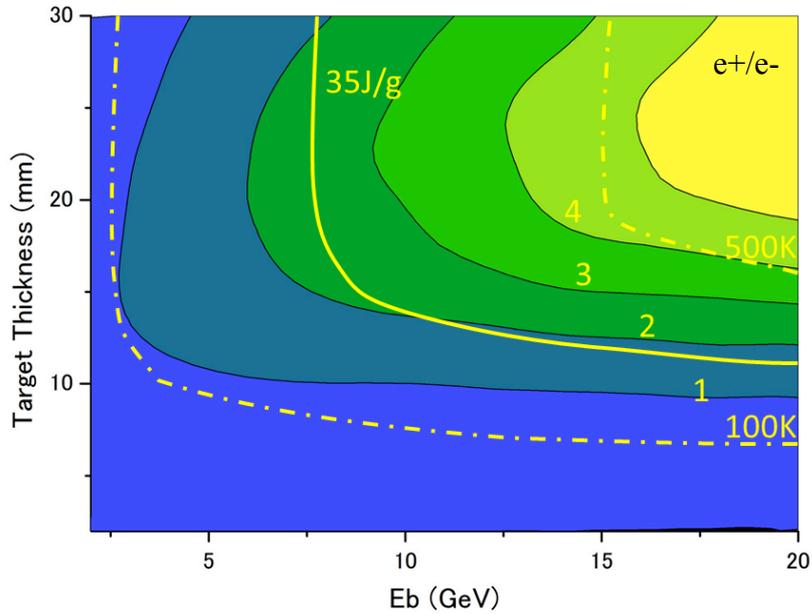

Figure 8. Number of accepted positrons per incident electron (colored contour) for a driving electron beam size of 4.0 mm. The lines for a PEDD of 35 J/g (solid line) and the peak temperature rise (dot dash line) are also shown. To estimate the PEDD and the peak temperature rise, the contributions of the 132 bunches in a triplet were accumulated; a bunch charge of 3.2 nC was assumed. If we assume that the target will be broken when the PEDD exceeds 35 J/g, the upper-right area above the 35 J/g line is excluded.

Figure 7 shows a typical distribution in the longitudinal phase space at the exit of the RF section after phase optimization. The damping ring acceptance limits the longitudinal phase to ±25MeV and ±3.46cm of the mean values as indicated in the Figure 7. Figure 8 shows a contour plot of the accepted number of positrons within the damping ring acceptance described above for a driving electron beam size of σ = 4.0 mm. Also shown in this figure are the estimated peak temperature rise and the line of the PEDD of 35 J/g which is the maximum tolerated value estimated by the previous study [5-7]. To determine the PEDD and the peak temperature rise, we assumed the 300 Hz scheme, 132 bunches in a triplet, and a bunch charge of 3.2 nC to accumulate the contributions to both PEDD and peak temperature rise.

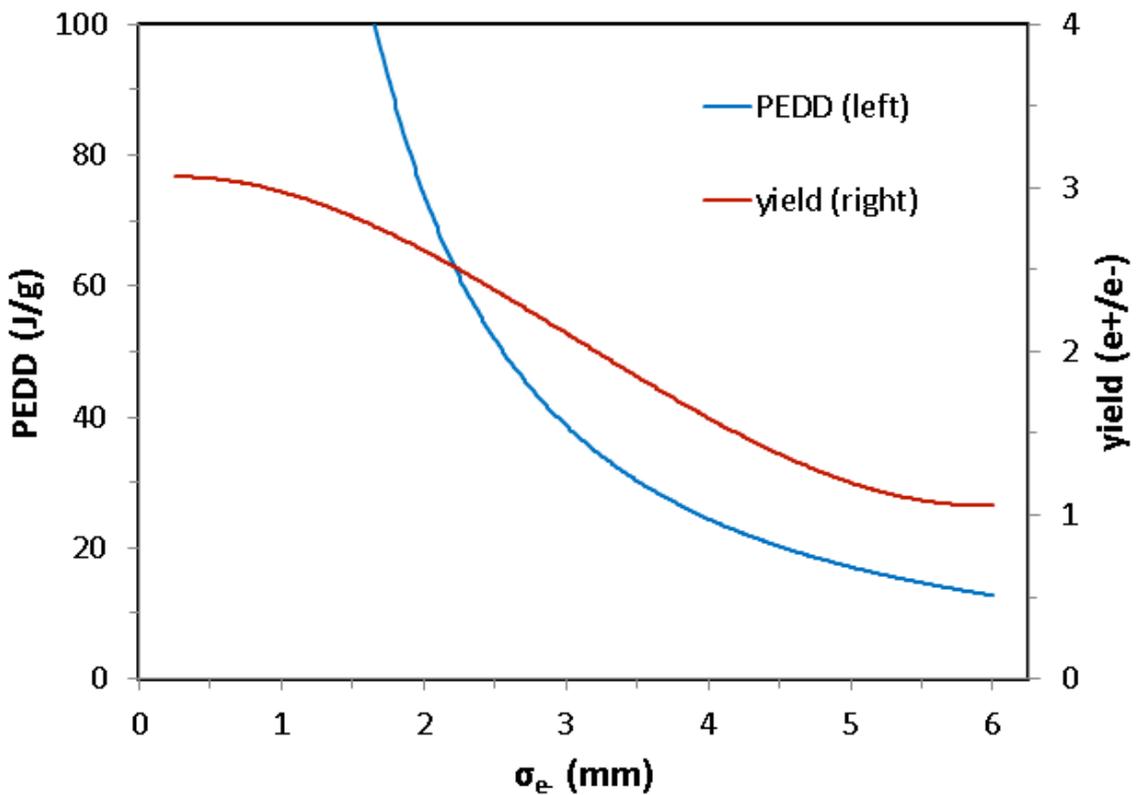

Figure 9. The PEDD (blue, left scale) and positron yield (red, right scale) as a function of drive electron beam size. The drive beam energy is 6 GeV, the bunch charge is 3.2 nC, and the target thickness is 14 mm. We assumed the 300 Hz scheme and 132 bunches in a triplet to accumulate the contributions.

Since the time duration which contributes to the target damage is uncertain, we estimated the potential damage due to spatial and temporal concentration of the energy deposit by assuming that all bunches in a triplet in 996 ns contribute equally.

As discussed in the previous section, the time duration of a triplet is shorter than the time of thermal diffusion. But the time between triplets (3.3 ms) is sufficiently long to achieve that each triplet hits a different position of the target rotating with tangential speed of about 5 m/s (see Figure 3a). Even if the tangential speed is lower than 5 m/s, the temperature rise can be acceptable. Figure 3d shows that a tangential speed of 0.5 m/s is acceptable concerning the temperature rise. With 0.5 m/s, the spatial separation of two successive triplets on the target is 1.7 mm which is smaller than the beam spot size if we employ a beam size larger than 1.7 mm in rms. Since thermal shock waves in the target develop within 1 ms or less, shock waves from successive triplets with 3.3 ms time interval should not be cumulative. If we use the PEDD as a measure of shock wave creation, a tangential speed of 0.5 m/s is acceptable also in this respect.

The actual choice of the tangential speed depends on further optimization and on the engineering design of the target system. Thus we need further studies for the final design of the positron source. The positron yield and the PEDD as function of the size of the driving beam are plotted in Figure 9. The parameters of the proposed positron source design are summarized in Table 1 assuming a 6 GeV drive beam of size 4 mm, which hits a tungsten target of 14 mm thickness. The average energies deposited in the target, AMD and RF sections are also shown in the table.

## 4. Conclusions

A conventional scheme is the only experienced scheme of positron generation in real accelerators. These experiences allow us to contain risk areas of positron generation in a target system. The drawback of the conventional positron source is that it has no capability to provide polarization. We need to replace the positron source when we need a polarized positron beam in future. Although the conventional scheme has such a drawback, it has a significant advantage in the viewpoint of a risk control which is vitally important in designing a very large system such as ILC. The conventional positron source proposed in this paper largely reduces the risk in target system by pulse stretching, generating about 2600 bunches of positrons by the 300 Hz drive electron beam with optimized beam and target parameters. It cures target thermal issues and enables us to employ a conventional positron generation scheme in the ILC. With respect to the risk control, the conventional scheme could be a suitable solution at the first stage of the ILC project.

Table 1. Parameters and results using a drive electron beam of 6 GeV with beam size 4.0mm, hitting a target of 14mm thickness.

| Parameters for target and captures | | Parameters for the 300Hz scheme | |
|---|---|---|---|
| Drive beam energy | 6 GeV | # drive e- /bunch | $2\times10^{10}$ |
| Beam size | 4.0 mm (rms) | # bunches/triplet | 132 (in 996 ns) |
| Target material | Tungsten | # bunches/train | 2640 (in 63 ms) |
| Target thickness | 14 mm | repetition of the trains | 5 Hz |
| Max. AMD field | 7 T | Results numbers in () are for the 300Hz scheme | |
| Taper parameter | 60.1/mm | e+ yield | 1.6 /e- |
| AMD length | 214 mm | PEDD in the target | 1.04 GeV/cm$^3$/e- (22.7 J/g) |
| Const. field | 0.5 T | Energy deposit in the target | 823 MeV/e- (35 kW) |
| Max. RF field | 25 MV/m | Energy deposit in the AMD | 780 MeV/e- (33 kW) |
| RF frequency | 1.3 GHz | Energy deposit in the RF section | 470 MeV/e- (20kW) |

**Acknowledgements**

We would like to appreciate Dr. M. Kuriki of Hirosima university for his valuable suggestions. We also would like to appreciate Dr. K. Yokoya of KEK, his critical comments were always useful to improve our ideas. Our heartfelt appreciation goes to Dr. L. Rinolfi of CERN and Dr. T. Kamitani of KEK for fruitful discussions. Dr. S. Guiducci of INFN/Frascati gave us constructive comments on the relation between the damping ring design and positron source. We would like to appreciate Dr. J. Rochford of CCLRC/RAL and Dr. I. Bailey of Cockcroft Institute for their help to evaluate target heat issues. We would like to thank valuable discussion with Dr. R. Chehab of University Lyon-1 and Dr. A. Variola of LAL. A part of this research received support of Global COE Program "the Physical Sciences Frontier", MEXT, Japan.